\documentclass[intlimits,twoside,a4paper]{article}
\usepackage[cp1251]{inputenc}
\usepackage[eqsecnum]{cmpj3}

\usepackage{bm}

\issue{2020}{23}{3}{33706}
\doinumber{10.5488/CMP.23.33706}
\title[Kernel polynomial method to Anderson transition in disordered $\beta$-graphyne]
{Kernel polynomial method to Anderson transition in disordered $\beta$-graphyne}
\author[G.X. Wang]{G.X. Wang}
\address{ School of Science, Henan Institute of Technology, 453003 XinXiang, China}
\date{Received February  11, 2020, in final form April 26, 2020}

\begin{document}

\maketitle

\begin{abstract}
By means of variable moment kernel polynomial method, we analyze the localization properties 
of $\beta$-graphyne sheet subjected to the Anderson disorder. To detect the localization
transition we focus on the scaling behavior of the normalized typical density of states. 
We find that there takes place a metal-insulator transition and the critical disorder strength is of the 
order of the bandwidth, which is contrary to the one-parameter scaling theory stating that for 
infinite two dimensional systems, all the electronic states are localized for an arbitrary strength of the
Anderson disorder. As its particular localization properties, it is reasonable to predict 
there will exist dc conductivity for $\beta$-graphyne at zero temperature.
\end{abstract}

\section{Introduction}

Graphyne, a single layer of carbon sheet, was first predicted by Baughman et al. in 
1987 as a layer consisting of $sp$ and $sp^2$ hybridized carbon atoms \cite{Baughman,Kang}. 
Graphyne is similar in structure to graphene but it is constructed by inserting carbon 
triple bonds ($-$C$\equiv$C$-$) into single bonds (C$-$C) in graphene. Although graphyne 
is topologically equivalent to graphene, the presence of triple bonds introduces rich 
optical and electronic properties \cite{Coluci} that are quite different from those of 
graphene, such as higher carrier mobility~\cite{Chen}, preferred chemical properties 
and strongly anisotropic optical adsorption \cite{Wang,Zheng}.  Because of these unique 
properties, graphyne has inspired researchers to explore its potential applications    
in electronics, catalysis, energy storage and biomedical fields \cite{Pei,Yue,Pan,Zhang,WangN,Liu}.

Due to different distributions of $sp$ and $sp^2$ hybridized carbon atoms, there were proposed various 
kinds of graphyne. Specially, four different types of graphyne 
were identified, namely $\alpha$-, $\beta$-, $\gamma$- and 6,6,12-graphyne \cite{Baughman,MalkoL}. 
In reference \cite{MalkoL}, it is reported that $\alpha$-, $\beta$- and 6,6,12-graphyne 
possess Dirac cone-like band structure around the Fermi level \cite{MalkoL}. Unlike 
graphene, the Dirac points of $\beta$-graphyne not only appear at high-symmetry points 
but also at low-symmetry points. According to these results, it is reasonable to expect 
that these graphynes can be considered as competitors to graphene and the study of graphyne 
allotropes will open insights into designing new-generation electronics.

So far, bulk graphyne sample is experimentally unobtainable but can be considered 
chemically stable. Tremendous efforts have been devoted to synthesizing and assembling a 
precursor and subunit of graphyne, a class of molecules known as 
dehydrobenzoannulenes (DBAs) \cite{HaleyMM,Kehoe}. Graphdiyne \cite{Haley}, 
which is a sub-structure of graphyne with benzene rings linked by diacetylene ($-$C$\equiv$C$-$C$\equiv$C$-$), 
was successfully synthesized on a copper substrate via a cross-coupling reaction 
using hexaethynylbenzene~\cite{LiGChem}. Moreover, graphdiyne can be fabricated by adopting low-temperature chemical 
vapor deposition \cite{Rong} or through a carbon-carbon coupling
reaction among constitutive monomers  at a liquid/liquid or gas/liquid interface~\cite{Matsuoka}. 
These exciting experimental progresses indicate that the synthesis of graphyne and its allotropes 
can be expected in the near future.

Since in real materials disorder is unavoidably present, it is of particular importance 
to understand how disorder affects the physics of electrons in the fabricated samples. 
Disorder can arise for various reasons, such as impurities (charged impurities, 
chemical impurities,  etc.), topological defects (vacancies, edge disorder, etc.), 
ripples or ad-atoms \cite{Querlioz,LiCT,Zare,Roche,Pogorelov}. For graphene, as the quality of 
samples is improving and synthesis techniques are developing, the dominant disorder varies 
from bulk disorder to edge defects. So far, researchers are devoted to fabrication of 
graphyne families and it is reasonable that bulk disorder will play a dominant role 
in these materials. When modelling a disorder, the Anderson model has been widely used 
in studying the basic localization features.

For graphene, due to the linear dispersion at the Dirac point, there was debated the issue whether the one-parameter scaling theory can hold \cite{Amini,Song,Bardarson,Fan}.
Amini et al. study the effect of on-site uncorrelated disorder on the electronic properties of graphene \cite{Amini}, 
and find that weak disorder can decrease the velocities of Dirac fermions, with extended states remaining.  
However, when the disorder strength increases to a large enough amount, there will be localized states around the Fermi point, 
which is consistent with the results of Naumis \cite{Naumis}. Subsequently, Schileede pointed out for an arbitrary disorder strength that
all the states are localized and there exists no mobility edge \cite{SchubertC}. Amini et al. acknowledged that the mobility edge might be induced by
the kernel \cite{Amini2}, but they argued the comment that the localization can occur throughout the whole spectrum because an arbitrary weak disorder strength 
is unreasonable since there exists minimal conductivity in graphene samples, and Anderson transition was observed by Bostwick et al. \cite{Bostwick}.
Recently, further study has been undertaken
\cite{Ardakani,Ilic,Fujimoto,Khademi,Lemut}.

Of late there exist plenty of well established numerical methods, such as 
exact diagonalization method, Monte Carlo simulations, Lanczos recursion method 
and kernel polynomial method (KPM). Every method is suited to particular problems 
and has severe limitations. For example, Lanczos algorithm is a powerful method to 
calculate extremal eigenstates of sparse matrices but it becomes unstable owing to the 
loss of orthogonality between the generated basis vectors. However, KPM, which expands 
the spectral function in terms of a complete set of orthogonal polynomials, can be 
stable and more convenient to deal with spectral densities and correlation functions. 
The spectral properties, such as LDOS, can be calculated to arbitrary precision without 
an exact diagonalization of the Hamiltonian \cite{Weibe}.

In this Letter, we consider $\beta$-graphyne sheet with the tight-binding approximation 
and focus on the effect of the Anderson-type bulk disorder \cite{Anderson}. With the 
modified KPM-the variable moment kernel polynomial method \cite{Schubert} (VMKPM), 
we calculate the local density of states (LDOS) and analyze the scaling behavior of 
normalized typical density of states (DOS). To detect the Anderson transition, the normalized 
typical DOS can serve as an order parameter.

\section{Model and method}

\begin{figure}[!t]
	\centerline{\includegraphics[width=0.38\textwidth]{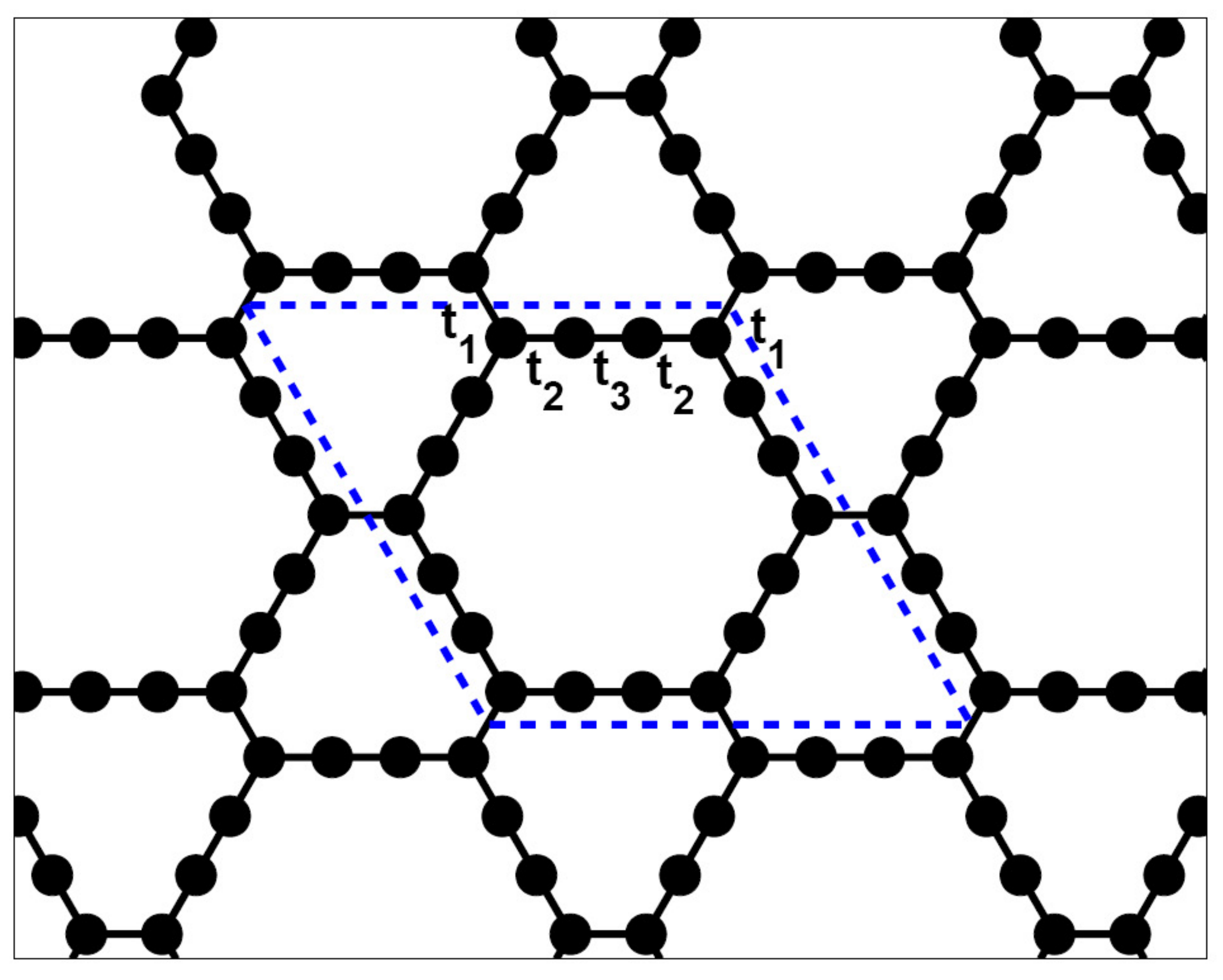}}
	\caption{(Colour online) The schematic diagram of lattice structure of
		$\beta$-graphyne. $t_1$, $t_2$ and $t_3$ denote the hopping
		amplitudes between the nearest-neighbor sites. The dashed lines
		specify the unit cell of $\beta$-graphyne.} 
	\label{fig-smp1}
\end{figure}

$\beta$-graphyne consists of a honeycomb lattice of carbon atoms with eighteen 
sublattices as shown in figure~\ref{fig-smp1}. The tight-binding Hamiltonian of $\beta$-graphyne 
with Anderson disorder can be written as 
\begin{equation}
{H} = \sum\limits_{\left\langle {ij} \right\rangle } { ( {t_{ij} c_i^{\dag} {c_j} + H.c.}  )}  + \sum\limits_{i  }  {{\epsilon _i}c_i^{\dag} {c_i}}\,,
\end{equation}
where   $c_{i}(c_i^{\dag})$ is the annihilation (creation) operator destructing 
(creating) an electron at  lattice site  $i$;  $\langle i, j\rangle$ represents the 
nearest-neighbor hopping with the hopping amplitude   $t_{ij}$, which is specifically 
denoted in figure~\ref{fig-smp1}; $\epsilon _i$  is the random on-site potential induced by Anderson 
disorder. Here, we assume that the random on-site potential $\epsilon _i$ has a uniform 
probability distribution as 
\begin{equation}
p( \epsilon _i ) = \frac{1}{\gamma }\theta \left( \frac{\gamma}{2} - \left| \epsilon _i \right|
\right),
\end{equation}
in which the parameter $\gamma$ is a measure of disorder strength and $\theta$ is Heaviside step function.

Many quantities can be used to describe localization properties, such as the averaged 
localization length, inverse participation number or the distribution of LDOS. Among 
these quantities,  LDOS is the most favorable to study Anderson localization both 
theoretically and experimentally with the fact that it can be applied to more complex 
situations and can be directly measured by scanning tunneling microscopy. The LDOS at lattice 
site  $i$ can be defined as
\begin{equation}
{\rho _i}\left( E \right) = \sum\limits_{n }  {{{\left| {\left\langle i \right|\left. n \right\rangle } \right|}^2}\delta \left( {E - {E_n}} \right)},
\end{equation}
where $\left | n \right\rangle$ is an eigenstate of $H$ with energy $E_n$ and $\left| i \right\rangle  = c_i^{\dag} \left| 0 \right\rangle $.

To distinguish a localized state from an extended one, it is convenient to compare two 
characteristic quantities: the typical DOS and the mean  DOS, which are defined, respectively, as
\begin{align}
{\rho _{\rm{ty}}}\left( E \right) &= \exp \left\lbrace 
{\frac{1}{{{K_r}{K_s}}}\sum\limits_{k = 1}^{{K_r}} {\sum\limits_{i =
			1}^{{K_s}} {\ln \left[ {\rho _i^k\left( E \right)} \right]} } }
\right\rbrace  ,\\
{\rho _{\rm{me}}}\left( E \right) &= \frac{1}{{{K_r}{K_s}}}\sum\limits_{k = 1}^{{K_r}} {\sum\limits_{i = 1}^{{K_s}} {\rho _i^k\left( E \right)} } ,
\end{align}
where $K_r$ denotes the number of disorder realizations and $K_s$ is the number of 
different lattice sites for a given realization. $\rho_{\rm{ty}}$ and $\rho_{\rm{me}}$ 
are both finite for extended
states while for localized states $\rho_{\rm{ty}}$ tends to zero but $\rho_{\rm{me}}$ 
remains finite.

To detect the localization transition  only for a single finite size system is 
insufficient \cite{SchubertC}. It is reasonable to take into account the scaling behavior 
of typical DOS, $\rho_{\rm{ty}}$. Consider the ratio of the typical and mean DOS
\begin{equation}
R\left( E \right)=\frac {\rho_{\rm{ty}} \left( E \right)}{\rho_{\rm{me}} \left( E \right)}.
\end{equation}
$R( E )$, the normalized typical DOS, is positive for extended states whereas it is equal to 
zero for localized states. Therefore, $R(E)$ can serve as an order parameter to reveal 
the position of the Anderson transition. By  calculating $R( E )$ for different 
disorder strength and lattice sizes, we can analyze the scaling behavior of the typical DOS.

Theoretically, LDOS can be evaluated to arbitrary precision using the KPM \cite{Weibe}, which 
is based on the expansion of the spectral function, such as $\rho_i( E )$, into
a finite series of Chebyshev polynomials $T_n( x ) = \cos ({n\arccos x} )$.  Since 
Chebyshev polynomials are defined in the interval [-1,1], correspondingly, the 
Hamiltonian $H$ needs to be rescaled before expansion with the rescaled 
Hamiltonian $\tilde{H}=(H - b)/a$ and rescaled energies $\tilde{E}= (E - b)/a$, 
where $a =( E_{\max } - E_{\min })/( {2 - \eta } )$ and $b = ( E_{\max } +E_{\min } )/2$. 
$\eta$ is a small positive number and it is not essential to have accurate bounds on the spectrum.

In terms of Chebyshev polynomials, LDOS can be approximated as
\begin{equation}
{\rho _i}( \tilde{E} ) = \frac{1}{\piup \sqrt {1 - {\tilde{E}^2}}}\left[ g_0{\mu _0} 
+ 2\sum\limits_{m = 1}^M {g_m {\mu _m}{T_m} ( \tilde{E}  )} \right],
\label{eq2.7}
\end{equation}
where the Chebyshev moments
%\begin{equation}
%\begin{split}
%{\mu _m} & = \int_{ - 1}^1 {{\rho _i} ( \tilde{E}  ){T_m} ( \tilde{E}  )\rd\tilde{E}}  \\
%& = \sum\limits_{n  }  {\left\langle i \right|\left. n \right\rangle \left\langle n \right|{T_m} ( {{\tilde{E}_n}}  )\left| i \right\rangle } \\
%& = \left\langle i \right|{T_m} ( \tilde{H} )\left| i \right\rangle .
%\end{split}
%\end{equation}
%
\begin{equation}
{\mu _m}  = \int_{ - 1}^1 {{\rho _i} ( \tilde{E}  ){T_m} ( \tilde{E}  )\rd\tilde{E}}  
 = \sum\limits_{n  }  {\left\langle i \right|\left. n \right\rangle \left\langle n \right|{T_m} ( {{\tilde{E}_n}}  )\left| i \right\rangle } 
= \left\langle i \right|{T_m} ( \tilde{H} )\left| i \right\rangle .
\end{equation}
Using the recursion relations between the Chebyshev polynomials, 
${T_{m + 1}}(\tilde{E}) = 2\tilde{E}{T_m}( \tilde{E} ) - {T_{m - 1}}( \tilde{E})$, these 
moments can be efficiently obtained  through a recursive procedure and the expression 
for $\mu_m$ can be very simple: $\mu_m=\left\langle i \right|\left. \alpha_m \right\rangle$ 
in which $\alpha_m= T_m ( \tilde{H} ) \left| i \right\rangle $. The most expensive 
computation is the multiplication between a sparse Hamiltonian matrix and a vector. 
If the matrix is computed \textit{on the fly}, without being stored, dimensions of the 
order can be carried out to $10^9$ or more. Since the Hamiltonian is a sparse matrix, 
the calculation can be done even on a desktop computer. In contrast to Lanczos, which 
is unstable because of loss of orthogonality and spurious degeneracy, Chebyshev iteration is completely stable.

Since the number of Chebyshev moments $\mu_m$ in equation~(\ref{eq2.7}) is finite, it is necessary 
to convolute the approximated function with kernel polynomials in order to damp out the 
Gibbs oscillations and ensure properties such as positivity and normalization. In this paper, 
we use the strictly positive Jackson kernel~\cite{Silver}
\begin{equation}
{g_{_m}} = \left( {1 - \frac{{m\phi }}{\piup }} \right)\cos \left( {m\phi } \right) + \frac{\phi }{\piup }\sin \left( {m\phi } \right)\cot \left( \phi  \right)
\end{equation}
with $\phi=\piup / \left( M+1 \right )$. The broadened peak for an order $M$ expansion 
of a $\delta$ function at position $x_0$ is a good approximation to a Gaussian of width $\sigma$ \cite{Weibe}
\begin{equation}
\sigma   = \sqrt {\frac{{M - x_0^2\left( {M - 1} \right)}}{{2\left( {M + 1} \right)}}\left( {1 - \cos 2\phi } \right)}
\approx \frac{\piup }{M}\sqrt {1 - x_0^2 + \frac{{4x_0^2 - 3}}{M}} .
\label{eq2.10}
\end{equation}
In order to distinguish between resolution and localization \cite{Schubert,SchubertG}, it 
is essential to ensure a uniform energy resolution, namely constant $\sigma$, over the 
whole spectrum by restricting the number of moments in equation~(\ref{eq2.10}).

\section{Calculation and discussion}

\begin{figure}[!t]
	\centerline{\includegraphics[width=0.40\textwidth]{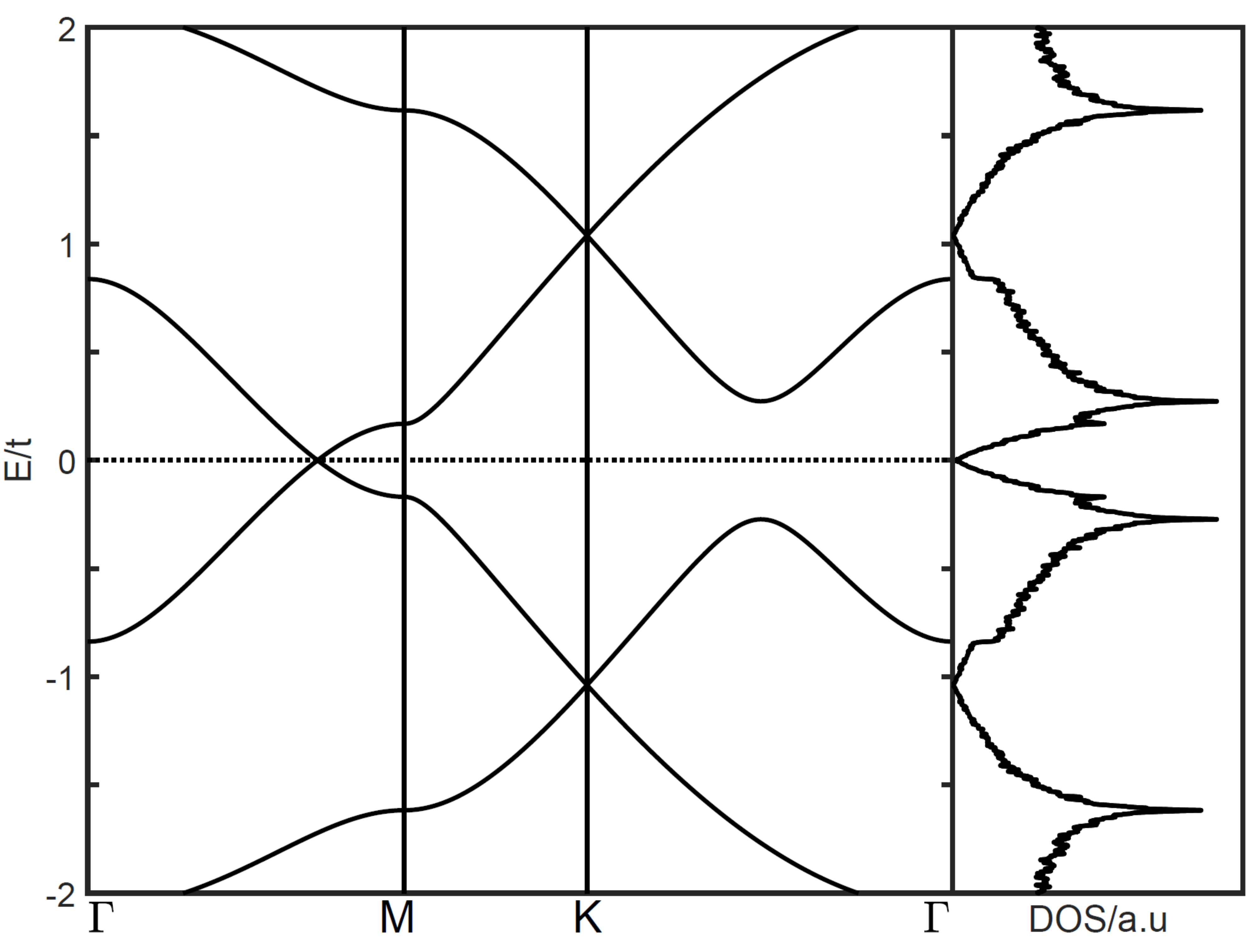}}
	\caption{ Band structure and DOS of ordered $\beta$-graphyne without expanding in Chebyshev
		polynomials.}
	\label{fig-smp2}
\end{figure}

In our calculation, the number of lattice sites is $N=4.5\times10^4-1.8\times 10^5$ with 
periodic boundary conditions. Following the reference \cite{LiuZ}, each carbon atom 
of $\beta$-graphyne is described by one 2$p_z$ orbital, and the corresponding hopping energies 
can be obtained, with $t_1=-2$ eV, $t_2=-2.7$ eV and $t_3=-4.3$~eV. In figure~\ref{fig-smp2}, we directly calculate
the band structure and its associated DOS of ordered $\beta$-graphyne without expanding in Chebyshev
polynomials, which is in consistence with the results of the  \cite{MalkoL}.

\begin{figure}[!t]
	\centerline{\includegraphics[width=0.50\textwidth]{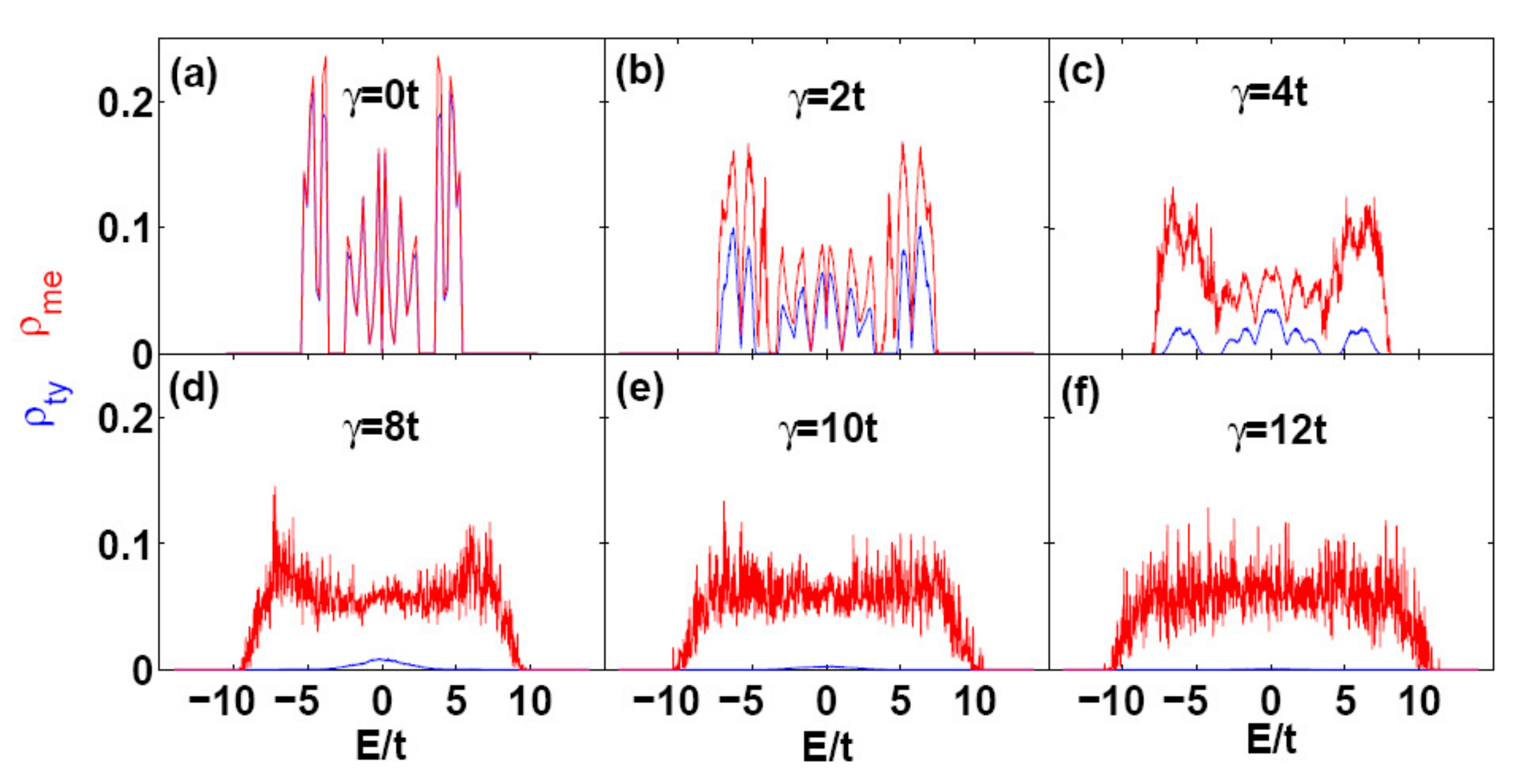}}
	\caption{(Colour online) Typical DOS (blue)  and mean DOS (red) for
		$\beta$-graphyne on a $1.8 \times 10^5$ lattice sites with periodic
		boundary conditions for different disorder strengths: (a) $\gamma=0$;
		(b) $\gamma=2t$; (c) $\gamma=4t$; (d) $\gamma=8t$; (e) $\gamma=10t$;
		and (d) $\gamma=12t$. Results are obtained using $N\sigma=45$ at $K_{\rm{s}}=32$ sites 
		for $K_{\rm{r}}=32$ realizations of disorder. Disorder strength and energy are in units of $t$ ($t=1$ eV).}
	\label{fig-smp3}
\end{figure}

We calculate the typical DOS $\rho_{\rm ty}$ and the mean DOS $\rho_{\rm me}$ of 
a $\beta$-graphyne sheet with $1.8\times 10^5$ lattice sites for different sites and 
disorder realizations $K=32 \times 32$. The results are shown in figure~\ref{fig-smp3}. As $\beta$-graphyne 
possesses the Dirac structure \cite{MalkoL}, the valence and conduction bands meet at the 
Dirac points and form the Dirac cone in the vicinity of the Dirac point. Consistently, 
the DOS is zero at the Fermi level, corresponding to $E=0$ in figure~\ref{fig-smp3} (a). 
For the ordered $\beta$-graphyne, $\rho_{\rm{me}} ( E )$ and $\rho_{\rm{ty}} ( E )$ coincide 
with each other, as shown in figure~\ref{fig-smp3} (a). With disorder increasing, starting from two boundaries 
of the spectrum, $\rho_{\rm{ty}} ( E )$ is suppressed and when $\gamma > 12t$ it vanishes, 
namely, at the critical disorder strength, the entire spectrum is localized. At the same time, 
the value of $\rho_{\rm{me}} (E)$ is always finite, no matter how disorder strength varies.

\begin{figure}[!t]
	\centerline{\includegraphics[width=0.50\textwidth]{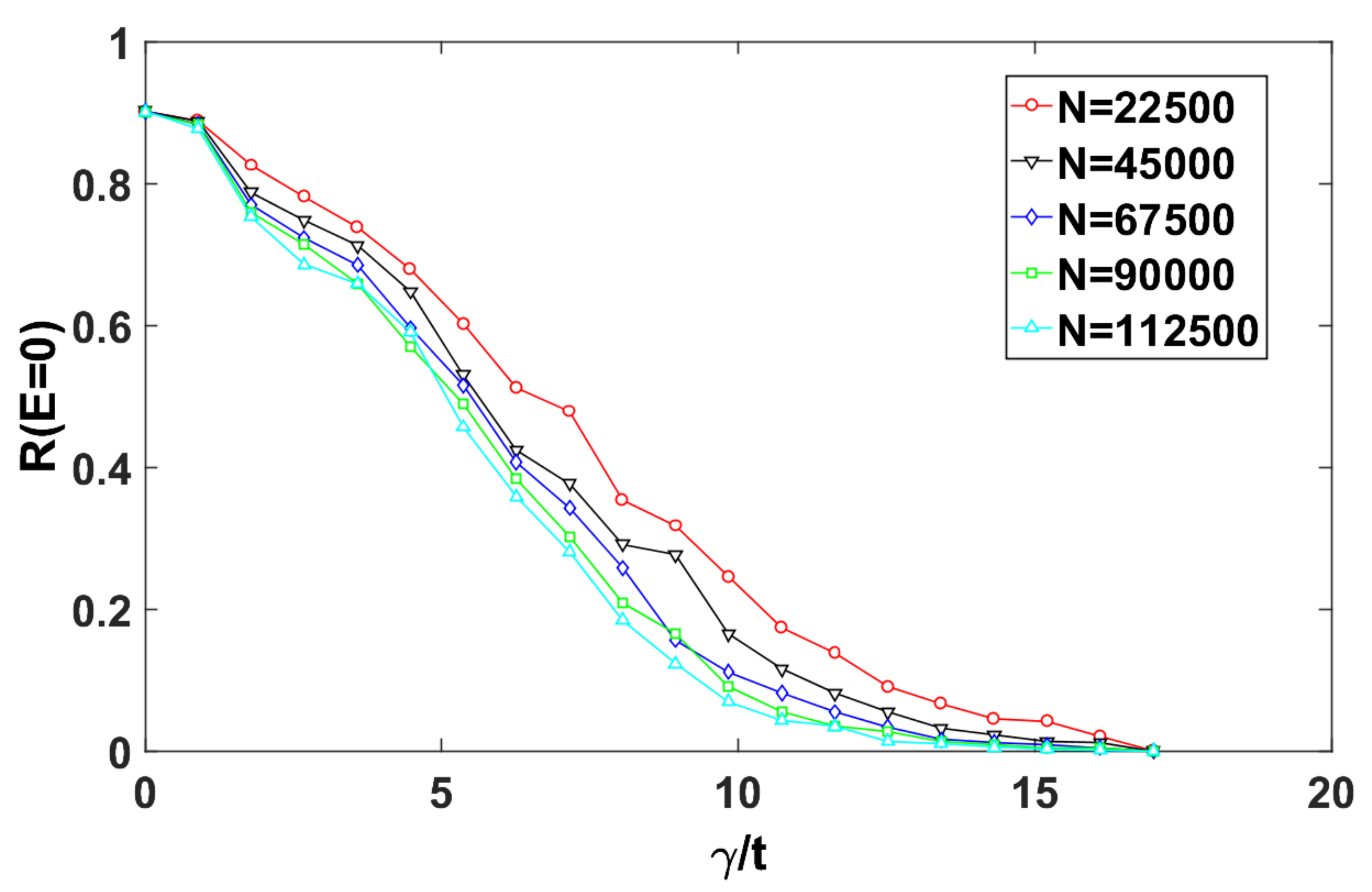}}
	\caption{(Colour online) Normalized typical DOS, $R(E)$, as a function of disorder,
		are calculated at $E=0$ for five different lattice sizes, with
		$K_{\rm{r}} \times K_{\rm{s}} =32 \times 32$ and $N\sigma=45$.}
	\label{fig-smp4}
\end{figure}

As mentioned above, only a single finite size system is insufficient to study the localization 
properties. We need to consider the normalized typical DOS,  $R( E )$. Figure~\ref{fig-smp4} presents $R( E )$ 
of five different system sizes for the energy $E=0$ as function of disorder strength. As Anderson 
transition can be influenced by the finite size of the system, and the resolution of KPM, for 
different system sizes $R( E )$, does not coincide very well but we can find that the curves tend to converge. 
More importantly, with an increasing disorder strength, $\rho_{\rm{ty}} \to 0$, which indicates Anderson 
localization. Then, by calculating $R( E )$, it is possible to reliably determine  the critical disorder 
strength $\gamma_c$. For example, we obtain the $W_{\rm{c}} \simeq 11t$ [$E=0$ and $R \left( E \right)=0.05$], 
which is of the order of bandwidth. The existence of critical disorder means that the localization behavior of $\beta$-graphyne 
is similar to 3D case, but contrary to the predicted 2D case of one-parameter theory.

\begin{figure}[!t]
\centerline{\includegraphics[width=0.55\textwidth]{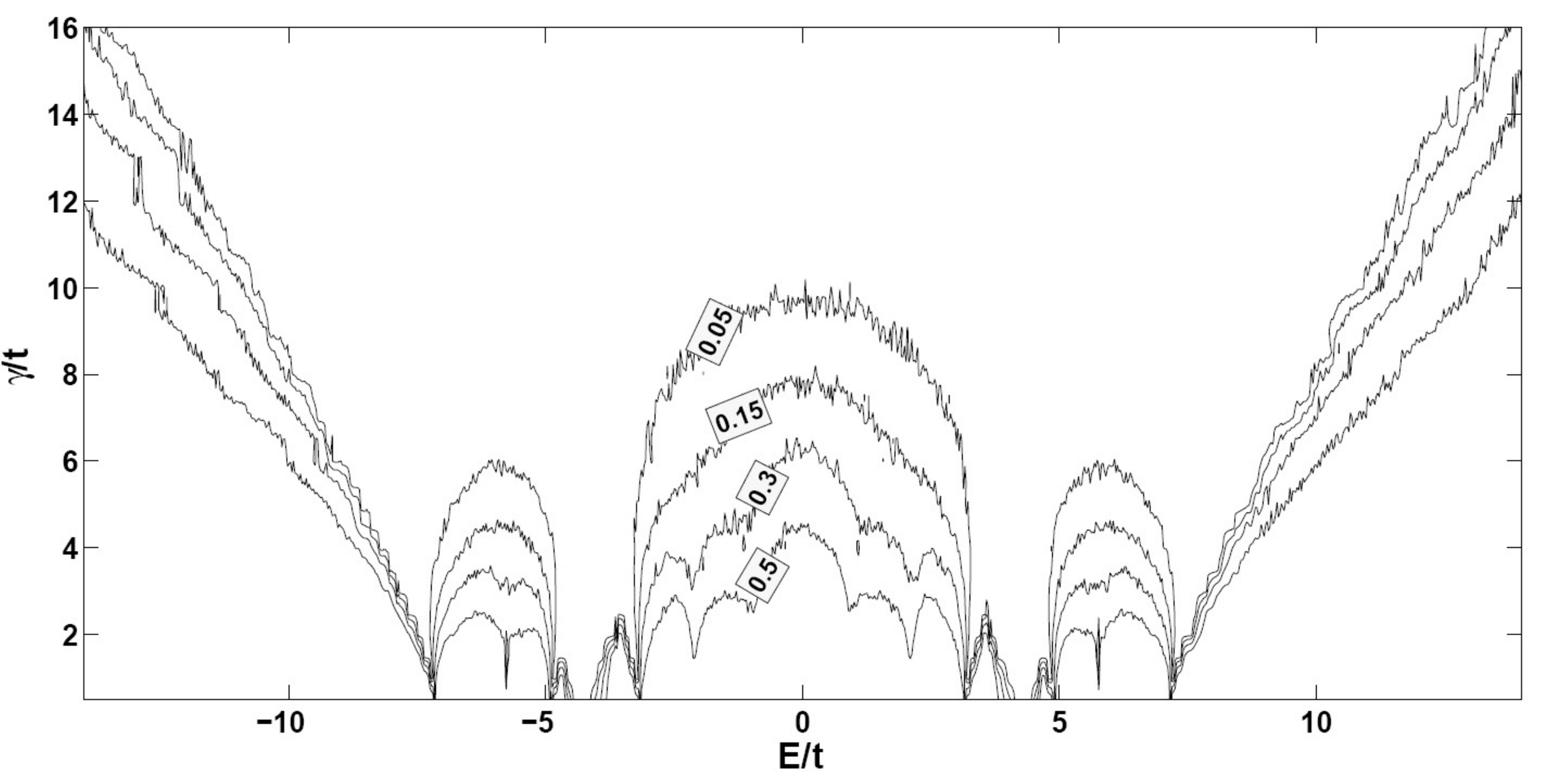}}
	\caption{ Contour plot of $R\left( E \right)$ in the energy-disorder plane. Mobility edge 
		obtained for $\rho_{\rm{ty}} / \rho_{\rm{me}} =0.05$ as well as the Lifshitz boundaries.}
	\label{fig-smp5}
\end{figure}

Figure~\ref{fig-smp5} shows the contours of the normalized typical DOS $R( E )$ in the energy-disorder plane. 
Since localized electrons do not contribute to the charge transport, the energy that separates 
localized states from the extended ones is called the mobility edge. From the contour map, we can 
find the mobility edge [$\rho_{\rm{ty}}( E )/\rho_{\rm{me}}( E ) = 0.05$] and apparently all 
states become localized for $\gamma >12t$. In figure~\ref{fig-smp5}, we can also find that the value of $R( E )$ 
at two boundaries of the spectrum is finite. However, from figure~\ref{fig-smp5} we know that $\rho_{\rm{me}} ( E )$ 
[which coincides with the standard density of states $\rho( E )$] is close to zero, which indicates that 
the probability of reaching these energy states is particularly small. Thus, these finite values 
of $R( E )$ will make no sense and the corresponding curve is known as the Lifshitz boundaries.

\begin{figure}[!t]
\centerline{\includegraphics[width=0.60\textwidth]{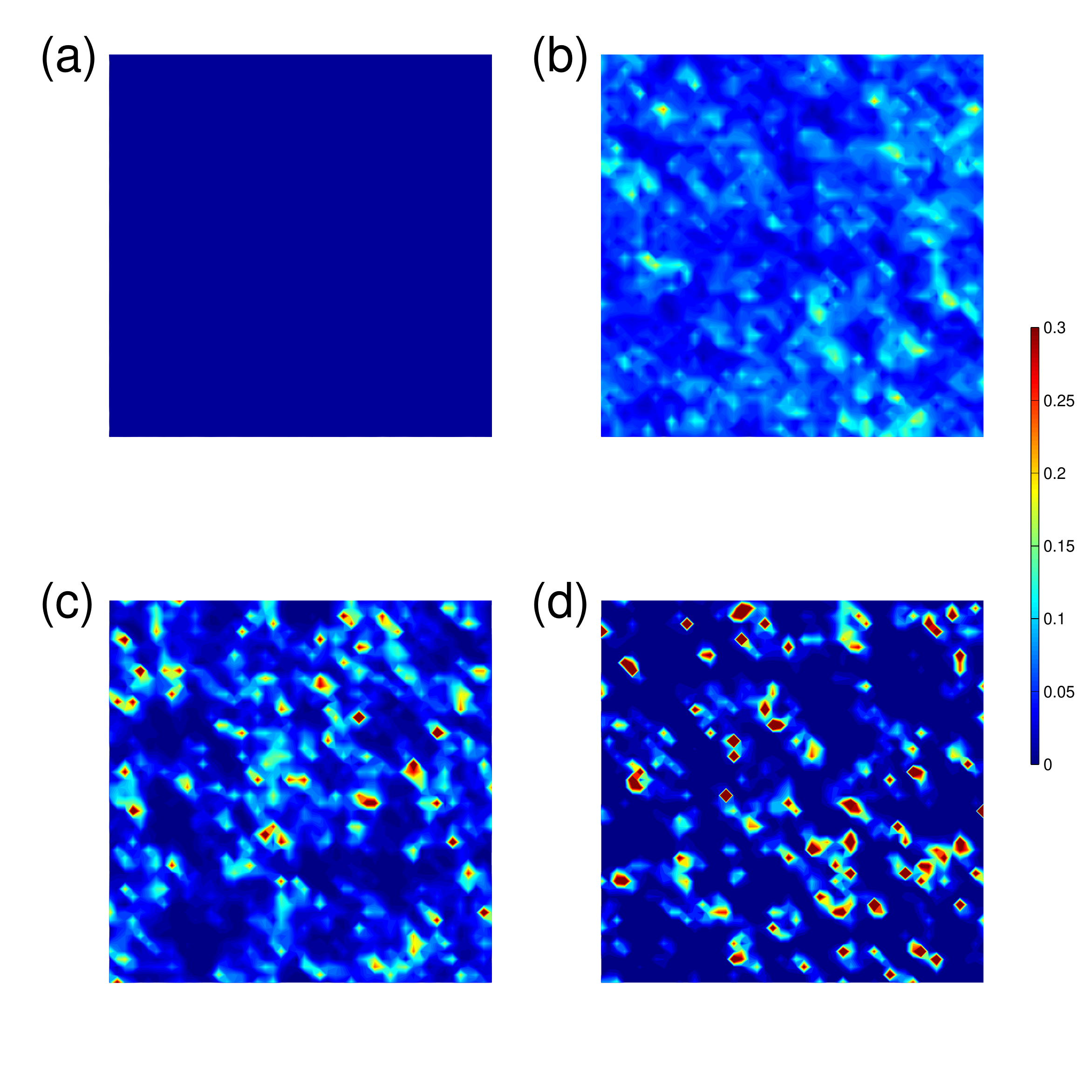}}
	\caption{(Colour online) LDOS map for different disorder strengths (a)
		$\gamma=0$; (b) $\gamma=4t$; (c) $\gamma=8t$; and (d) $\gamma=12t$ at energy $E=0$.
		Here, the calculations were performed on  $4.5\times 10^4$ lattice
		sites  with $N\sigma=45$.}
	\label{fig-smp6}
\end{figure}

For a given energy $E$, $\rho_i ( E )$ can measure the local amplitude of the wave function 
at site $i$.  In order to understand the internal structure of the extended and 
localized states, we calculate the LDOS in the band center ($E=0$) for one realization 
of disorder in figure~\ref{fig-smp6}. For the ordered $\beta$-graphyne sheet, the distribution of LDOS 
is uniform [figure~\ref{fig-smp6} (a)]. When the disorder strength $\gamma=4t$, from figure~\ref{fig-smp6} (b) we can see that
although the distribution of  LDOS becomes nonuniform, the states  still remain delocalized. 
As the disorder strength increases to $\gamma=8t$, the states clump into some finite regions 
of clusters, as shown in figure~\ref{fig-smp6} (c), which indicates that the localization takes place. However, between 
the clusters, there exist bridges where the LDOS is very small but nonvanishing. When the 
disorder strength reaches $\gamma=12t$, the states are confined to many isolated islands, 
which implies that these states are totally localized and the metal turns into an insulator, 
that is to say, the Anderson transition occurs.

\section{Conclusion}

In summary, we use typical DOS $\rho_{\rm{ty}}$ and the normalized typical DOS $R( E )$, 
which can be efficiently and precisely calculated by means of variable moment kernel 
polynomial method, in order to detect the localization transition of disordered $\beta$-graphyne sheet. 
When modelling a disorder, the Anderson model is widely used in studying the basic localization features. 
One parameter scaling theory predicts that disorder can induce localization transition in 3D systems while 
all states are localized for infinite 2D case. However, the metal-insulator transition in hydrogenated graphene \cite{Bostwick}, 
disordered GNRs \cite{Adam}, and Si-MOSFET inversion layers \cite{Kravchenko} have been reported, which are 
out of reach for Anderson model. For a single layer of disordered $\beta$-graphyne sheet, we have proved that 
it also exhibits metal-insulator transition and the critical disorder strength is of the order of the bandwidth, 
which might be explained by percolation-based approaches \cite{Adam,Sarma}. In our work, we only numerically analyze
the localization properties of $\beta$-graphyne. More details, such as the localization length, conductivity \cite{Fan,FanZ},
the contribution of Dirac nodes and the reduced back-scattering, and ``resilience'' properties of the Dirac nodes against 
weak to moderate disorder \cite{Amini,Castro}, need further investigation. Furthermore, we doubt that other allotropes, such as $\alpha$- and 6,6,12-graphyne, 
may possess similar localization properties, which will be investigated systematically later.

\section{Acknowledgements}
The work is supported by the Doctoral Program Foundation of Henan Institute of Technology.

\newpage
\ukrainianpart

\title{Метод поліномного ядра для андерсонівського переходу  у невпорядкованому  $\beta$-графіні}
\author{Г.К. Ванг}
\address{Школа природничих наук, Хенанський технологічний інститут, 453003 Сіньсян, Китай}
\makeukrtitle

\begin{abstract}
\tolerance=3000%
З допомогою методу поліномного ядра зі змінним моментом проаналізовано локалізаційні властивості листа $\beta$-графіна під впливом 
андерсонівського безладу. Для виявлення локалізаційного переходу ми зосередилися на поведінці скейлінгу нормалізованої типової густини станів. 
Встановлено існування переходу метал-ізолятор, причому сила критичного безладу є порядку ширини зони, що протирічить однопараметричній теорії 
скейлінгу, яка стверджує, що для нескінченних двовимірних систем усі електронні стани є локалізованими при довільній силі андерсонівського безладу. 
Стосовно конкретних локалізаційних властивостей, можна передбачити, що спостерігатиметься провідність постійного струму для $\beta$-графіну 
при нульовій температурі.
\keywords $\beta$-графін, метод поліномного ядра, андерсонівський безлад, локалізаційні властивості
\end{abstract}

\lastpage
\end{document}